# The Spearman-Brown Formula and Reliabilities of Random Test Forms

Jules L. Ellis

Faculty of Psychology, Open University of The Netherlands

Behavioural Science Institute, Radboud University Nijmegen

**Author note**

Jules L. Ellis https://orcid.org/0000-0002-4429-5475

Jules L. Ellis is now at Open University of The Netherlands.

Correspondence should be addressed to Jules L. Ellis, Faculty of Psychology, Open University of The Netherlands, Heerlen, The Netherlands. E-mail: jules.ellis@ou.nl

**Abstract**

It is shown that the psychometric test reliability, based on any true-score model with randomly sampled items and conditionally independent errors, converges to 1 as the test length goes to infinity, assuming some fairly general regularity conditions. The asymptotic rate of convergence is given by the Spearman-Brown formula, and for this it is not needed that the items are parallel, or latent unidimensional, or even finite dimensional. Simulations with the 2-parameter logistic item response theory model reveal that there can be a positive bias in the reliability of short multidimensional tests, meaning that applying the Spearman-Brown formula in these cases would lead to overprediction of the reliability that will result from lengthening the tests. For short unidimensional tests under the 2-parameter logistic model the reliabilities are almost unbiased, meaning that application of the Spearman-Brown formula in these cases leads to predictions that are approximately unbiased.

Keywords: reliability, generalizability, item response theory, Spearman-Brown, random parallel, convergence, bias.

**The Spearman-Brown Formula and Reliabilities of Random Test Forms**

The (generalized) Spearman-Brown formula (after Brown, 1910; Spearman, 1910) states that if a test with reliability $\rho_1$ is lengthened with factor $n$, the reliability of the lengthened test will be

$$\rho_n = \frac{n\rho_1}{1 + (n-1)\rho_1}$$

This formula is treated in many psychometric textbooks as an aid to "prophesize" what the reliability of a test would be after changing its length (e.g., Allen & Yen, 1979, p. 85-88; Furr & Bacharach, 2008, p. 127; Nunnally, 1978, p. 243; Reynolds & Livingston, 2012, p. 131; Webb, Shavelson & Haertel, 2006). It has also been used in studying how the power of statistical tests depends on the test length (Feldt, 2011, p. 425; Ellis, 2013, p. 19). Many textbooks mention the caveat that the formula is based on the assumption that the test components (i.e. items or groups of items) are parallel (e.g., Lord & Novick, 1968, p. 139), which means that the test components would have identical true scores and error variances, an assumption that is usually not realistic. Is it possible to relax this assumption? Obviously, as many psychometricians will be aware of, it is also sufficient if the test components are parallel up to an additive constant, since adding a constant to one component would affect neither $\rho_1$ nor $\rho_n$. Still, the question remains whether it is possible to relax this assumption.

Lord (1955) and Cronbach et al. (1972) argue that the assumption of parallel items is not needed to estimate reliability if one assumes that the items are randomly drawn from a large pool. It can be argued that this is also true for the Spearman-Brown formula. This article investigates the matter from the perspective of item response theory (IRT). It will be assumed that the test maker starts with a test of given length, for example 10 items, and that the reliability of the test, as computed from the IRT model, is correctly known to the test maker. The test maker wants to predict what the reliability will become after changing the test length to, for example 25 items, randomly drawn from the same pool. The research question in this article is under which circumstances the Spearman-Brown formula can be used for such predictions. The question is more specifically whether the accuracy of the Spearman-Brown formula depends on the dimensionality of the items, as is suggested by the warning that the items need to be parallel, or that random sampling of items is sufficient to assure its accuracy even if the items are multidimensional, as seems to be suggested in generalizability theory.

The next section reviews the role of the Spearman-Brown formula in generalizability theory. It will be argued that the Spearman-Brown formula seems to be used implicitly in some applications of generalizability theory, but that its validity does not obviously extend to short tests and IRT-based reliabilities. After introducing detailed notation and basic definitions in a separate section, a theorem will be presented, stating that if the test length goes to infinity, the Spearman-Brown formula holds approximately for IRT-based reliabilities of test forms consisting of randomly selected items. Following this, simulation studies will assess how good the approximation is for short test forms of randomly selected items that obey a one-dimensional or multidimensional two-parameter logistic (2PL) IRT model with normally distributed latent traits.

## The Spearman-Brown Formula in Generalizability Theory

In generalizability theory (Cronbach et al., 1972; Gleser, Cronbach & Rajaratnam, 1965; Webb, Shavelson & Haertel, 2006) it is assumed that the items are randomly sampled, and the assumption of parallel test items is often considered unnecessary. For example, Rajaratnam, Cronbach & Gleser (1965, p. 40) discuss the concept of "randomly parallel tests", which "are formed by drawing items randomly from the universe as a whole", and they state "For randomly parallel tests, $\alpha$ obeys the Spearman-Brown formula as *k'/k* departs from 1.00" (p. 50). A version of the Spearman-Brown formula is often used implicitly in generalizability theory. For example, consider a one-facet design of items and denote the observed-score variance, universe-score variance, residual variance, and generalizability for a test of length $n$ with $\sigma_n^2$, $\tau^2$, $\varepsilon_n^2$, and $\rho_n$ respectively. Assuming $\sigma_n^2 = \tau^2 + \varepsilon_n^2$ and defining $\rho_n = \tau^2/\sigma_n^2$, *the Spearman-Brown formula holds for $\rho_n$ if and only if* $\varepsilon_n^2 = \varepsilon_1^2/n$. The identity $\varepsilon_n^2 = \varepsilon_1^2/n$, or versions of it for more complex designs, is routinely assumed in studies on the optimization of generalizability (e.g., Marcoulides, 1993, 1995, 1997; Marcoulides & Goldstein, 1990, 1992; Meyer, Liu & Mashburn, 2014; Peng, Li & Wan, 2012; Sanders, 1992; Sanders et al., 1989, 1991; Woodward & Joe, 1973). For example, Marcoulides (1993) states that in a one-facet study, the variance of the sample mean can be partitioned as (in adapted notation)

$$\sigma_{\bar{X}}^2 = \frac{\sigma_{\text{persons}}^2}{n_{\text{persons}}} + \frac{\sigma_{\text{items}}^2}{n_{\text{items}}} + \frac{\sigma_{\text{residual}}^2}{n_{\text{persons}} n_{\text{items}}}$$

Similar formulas are used by the other authors cited above, and these formulas serve a similar role as the Spearman-Brown formula in that they predict the effect of changing the number of observations.

It is important to note that the above formula takes into account both sampling of persons and sampling of items, which implies that the variance $\sigma_{\bar{X}}^2$ is taken over all possible random test forms, and $\sigma_{\text{residual}}^2$ is taken over all items of the universe. If only the scores from a single finite random test form are considered, this equation will generally not hold exactly with the same values of the variance components. That is, if a single random test form is denoted by $S_n$, then $\text{var}(\bar{X}|S_n) \neq \sigma_{\bar{X}}^2$ and $\text{var}(residual|S_n) \neq \sigma_{\text{residual}}^2$, in general. If a test maker nevertheless uses the optimization procedure derived from the formula for $\sigma_{\bar{X}}^2$, and applies it to a single test form $S_n$, then the test maker seems to assume implicitly that the variance components are constant across all test forms (e.g., $\text{var}(residual|S_n) = \sigma_{\text{residual}}^2$). This assumption is not necessarily valid for non-parallel items.

Similarly, Cronbach et al. (1972) define a generalizability coefficient that they denote as $E\rho^2$, and that can be estimated by Cronbach's coefficient alpha, as $E\rho^2 := \sigma_{\text{persons}}^2 / (\sigma_{\text{persons}}^2 + \sigma_{\text{residual}}^2/n)$. However, it should be note that the usage of the expectation symbol is merely a suggestive notation, and that there is not literally an expectation being taken (Cronbach et al., 1972, p. 82; Brennan, 2001, p. 35). In the present article it would be rather confusing to use the expectation symbol for something that is not an expectation, and therefore we will rather use the symbol $\rho_n := \sigma_{\text{persons}}^2 / (\sigma_{\text{persons}}^2 + \sigma_{\text{residual}}^2/n)$ (note the omission of the square sign). From this definition it follows immediately that $\rho_n$ obeys the generalized Spearman-Brown formula, as pointed out above. However, this does not mean that an estimated reliability $\hat{\rho}_n$ based on one random parallel form will strictly obey the Spearman-Brown formula. It does not even mean that the expected value $\mathbb{E}\hat{\rho}_n$ (where now $\mathbb{E}$ actually denotes expectation) obeys it. Nevertheless, if $\hat{\rho}_n$ is a consistent estimate of $\rho_n$, then $\hat{\rho}_n$ must approximately obey the Spearman-Brown formula for large $n$ because $\rho_n$ obeys it. The question of this article is: does this also work for reliability coefficients that are not explicitly designed to work in generalizability theory but, for example, rather in a latent variable framework, such as McDonalds omega (see Zinbarg, Revelle, Yovel, & Li, 2005; Zinbarg, Yovel, Revelle, & McDonald, 2006) or the 'empirical reliability' in IRT models (e.g., Kim and Feldt, 2010)? And how well does it work for short tests (small $n$)?

These questions require the usage of both IRT and generalizability theory. The present approach to combine generalizability theory with IRT will be different from the one of Briggs and Wilson (2007) and Glas (2012). These authors use a parametric (logistic or normal ogive) model, and the variance components in their generalizability theory pertains to these latent parameters. In contrast, the formalization of the next section will not require a specific parametric model, and different parametrizations of the same model will entail the same variance components. It is also closer to the usual application of generalizability theory, applied to the observed scores.

## Notation and General Assumptions

### Notation of Conditional Expectations

This article will use results from the measure-theoretical foundations of conditional expectation (e.g. Austin & Pachenko, 2014; Billingsley, 1986; Majerek et al., 2005), and in order to do so smoothly, it will use the notation that is common in the theory of conditional expectations, which differs from the conventional notation in generalizability theory. Suppose $\Gamma$ is a set of persons and $\Omega$ is a set of test items and $X$ is a function on $\Gamma \times \Omega$. Let $V$ be a randomly selected person and let $R$ be a randomly selected item. The conditional expectation of $X$ given the person will be denoted as $\mathbb{E}(X|V)$. This is a random variable, a function of random variable $V$, and for a specific person $p$ this random variable assumes the value $\mathbb{E}(X|V=p)$. Many authors in generalizability theory would rather write $\mathbb{E}_i(X_{pi})$ for this. Similarly, the expectation over both persons and items will be written as $\mathbb{E}(X)$, not $\mathbb{E}_p\mathbb{E}_i(X_{pi})$.

**Definition of Item True scores and Item Error Scores**

In terms of generalizability theory we will assume a universe of admissible observations that consists of one facet, the items. Assume that we have a set of test items $\Omega$ where individual items are indicated by subscript $\omega \in \Omega$. Assume that each item $\omega \in \Omega$ has an observed-score random variable $X_\omega$ with finite non-zero variance.

Assume that there is some σ-field $\mathcal{F}$ for which we define $T_\omega = \mathbb{E}(X_\omega|\mathcal{F})$ and $E_\omega = X_\omega - T_\omega$ for each item $\omega \in \Omega$. Assuming these conditional expectation exist, this implies that $\mathbb{E}(E_\omega) = 0$ and $\text{cov}(E_\omega, T_\upsilon) = 0$ for all $\omega, \upsilon \in \Omega$. We assume furthermore that

$$\{E_\omega | \omega \in \Omega\} \text{ is conditionally independent given } \mathcal{F}$$

which implies that $\text{cov}(E_\omega, E_\upsilon)$ for all $\omega, \upsilon \in \Omega, \omega \neq \upsilon$. The $T_\omega$ will henceforth be called the true-score variables, the $E_\omega$ will be called the error-score variables, but it should be emphasized that we do not assume that these variables necessarily have the interpretation that is often given to them in texts on classical test theory (e.g., Lord & Novick, 1968). For example, we do not say that $T_\omega$ should be defined by an infinite series of replications within subjects. A few examples:

1. Assume that the items satisfy a latent variable model, as defined by Holland and Rosenbaum (1986), and that the $X_\omega$ are conditionally independent given some latent variable vector $\boldsymbol{\Theta}$. Note that we do not put any restriction on the dimensionality of $\boldsymbol{\Theta}$. Following Lord (1980, p. 46), Stout (1990), and Kim and Feldt (2010, p. 180), among others, we can use this to define 'model-based' true-score variables $T_\omega = \mathbb{E}(X_\omega|\boldsymbol{\Theta})$, and corresponding error-score variables $E_\omega = X_\omega - T_\omega$. This is a special case of the above definition if we set $\mathcal{F}$ equal to the σ-field generated by **Θ**. A comparable definition is given by Zimmerman (1976), albeit without explicit reference to a latent variable. The assumption that the $\{E_\omega | \omega \in \Omega\}$ are conditionally independent given $\mathcal{F}$ now corresponds to the assumption of "local independence" of item response theory.

2. Ellis and Junker (1997; also Junker & Ellis, 1997) argue that one can alternatively define 'tail-conditional' true-score variables $T_\omega = \mathbb{E}(X_\omega | \boldsymbol{\tau}(\mathbf{X}))$, where **τ(X)** is the tail-sigma field of the observed-score variables. This is a special case of the above definition if we set $\mathcal{F}$ equal to **τ(X)**.

3. Assume that the items satisfy a linear or nonlinear regression model with a set of predictors **U**. The predicted scores are $T_\omega = \mathbb{E}(X_\omega | \mathbf{U})$, and the residuals are $E_\omega = X_\omega - T_\omega$. This is a special case of the above definition if we set $\mathcal{F}$ equal to the σ-field generated by **U**. The assumption that the $\{E_\omega | \omega \in \Omega\}$ are conditionally independent given $\mathcal{F}$ now corresponds to the assumption that the residuals are conditionally independent given the predictors.

4. If sampling of observed scores within persons is defined, as described in Lord and Novick (1968, Chapter 2), and $V$ is a variable that indicates the persons, then let $T_\omega = \mathbb{E}(X_\omega | V)$, and $E_\omega = X_\omega - T_\omega$. This is a special case of the above definition if we set $\mathcal{F}$ equal to the σ-field generated by $V$. The assumption that the $\{E_\omega | \omega \in \Omega\}$ are conditionally independent given $\mathcal{F}$ means that the error-score variables are independent within each person, which is called "experimental independence" by Lord and Novick (1968).

Note that the error-score variables are not necessarily independent, even though their covariance is zero and they are conditionally independent given $\mathcal{F}$. Assume that the aforementioned variables have finite and positive variance. Let the standard deviations of $X_\omega, T_\omega$ and $E_\omega$ be $\sigma(\omega), \tau(\omega)$ and $\varepsilon(\omega)$, respectively.

**Assumption of Random Selection of Items**

Similar to Hunter's (1968) probabilistic foundation of generalizability theory, we will assume that some probability measure is defined for $\Omega$, and that $R_1, R_2, \ldots$ is an infinite sequence of independent identically distributed (i.i.d.) random variables with range in $\Omega$. Here, $R_i$ is supposed to be the name or number of the *i*-th test item during the random selection. Consistent with large parts of generalizability theory (e.g., Cronbach et al., 1972), it will be assumed in the sequel that the item pool is infinitely large and that it is almost impossible that the same the item is included twice, i.e. $P(R_i = R_j) = 0$ for $i \neq j$; $i, j \in \mathbb{N}$. It will furthermore be assumed that the $R_1, R_2, \ldots$ are independent of $\mathcal{F}, X_\omega, T_\omega$ and $E_\omega$ for all $\omega \in \Omega$.

**Assumptions on Moments**

It was already assumed that $0 < \mathrm{var}(X_\omega) < \infty$, $\mathrm{var}(T_\omega) < \infty$, and $\mathrm{var}(E_\omega) < \infty$ for all $\omega \in \Omega$, and it will furthermore be assumed that $\mathrm{var}(\varepsilon^2(R_i)) < \infty$ and $\mathrm{var}(\tau^2(R_i)) < \infty$ and $|\mathbb{E}(T_{R_i})| < \infty$ and $\mathbb{E}(T_{R_i}^2) < \infty$ for all $i \in \mathbb{N}$.

**Definition of Test True Scores and Universe Scores**

The test length will be denoted by $n \in \mathbb{N}$. Let $\boldsymbol{S}_n := (R_1, R_2, \ldots, R_n)$; this is a random vector, and each realization of it is a random test forms of length $n$. The observed-score variables of the random test form are

$$X_{R_1}, X_{R_2}, \ldots, X_{R_n}$$

These are different from the original observed-score variables in that the items are shuffled. For example, if $A, B \in \Omega$, then $X_{R_1} = X_A$ for some subjects, but $X_{R_1} = X_B$ for some other subjects. For random test form $\boldsymbol{S}_n$, define the test observed-score, true-score, and error-score variables as

$$X_{\boldsymbol{S}_n} := \sum_{i=1}^{n} X_{R_i}/n$$

$$T_{S_n} := \sum_{i=1}^{n} T_{R_i}/n$$

$$E_{S_n} := \sum_{i=1}^{n} E_{R_i}/n$$

Since it is assumed that $|\mathbb{E}(T_{R_i})| < \infty$, we can define

$$T_\infty := \mathbb{E}(T_{R_1}|\Theta),$$

which can be viewed as the model-based universe-score variable.

**Assumptions on Moments and Definition of Reliability**

It was already assumed that $\text{var}(X_\omega) < \infty$, $\text{var}(T_\omega) < \infty$, and $\text{var}(E_\omega) < \infty$ for all $\omega \in \Omega$, and it will furthermore be assumed that $\text{var}(\varepsilon^2(R_i)) < \infty$ and $\text{var}(\tau^2(R_i)) < \infty$ and $|\mathbb{E}(T_{R_i})| < \infty$ and $\mathbb{E}(T_{R_i}^2) < \infty$ for all $i \in \mathbb{N}$. Define the *reliability of test form* $S_n$ as

$$\rho(S_n) := \frac{\text{var}(T_{S_n}|S_n)}{\text{var}(X_{S_n}|S_n)}$$

For example, if the item pool is $\Omega = \{A, B, C, \dots\}$ and $n = 3$ then one random test form could be $S_n = (A, D, C)$ with reliability $\rho((A, D, C))$.

It can easily be shown that the Spearman-Brown formula holds for a different form of reliability. A sequence of variables $Y_1, Y_2, \dots$ is called exchangeable if for every $n \in \mathbb{N}$, every subsequence of $n$ variables from $Y_1, Y_2, \dots$ has the same joint distribution. The original variables $X_\omega$ are not exchangeable in general; for example, items in the Rasch model with different difficulty parameters have different means and can therefore not be exchangeable. The 'shuffled' variables $X_{R_1}, X_{R_2}, \dots$ are exchangeable nonetheless, because the variables $R_1, R_2, \dots$ are i.i.d. According to the de Finetti-Hewitt-Savage theorem (e.g., Austin & Pachenko, 2014, theorem 3), exchangeability implies that the variables $X_{R_i}$ are conditionally i.i.d. given their empirical distribution function.

Zimmerman (1976, theorem 2) showed that conditional i.i.d. variables satisfy a form of the Spearman-Brown formula. Although Zimmerman did not discuss randomly shuffled variables, his result can be applied here. Let $\boldsymbol{S}'_n = (R_{n+1}, R_{n+2}, \ldots, R_{2n})$ and $X_{\boldsymbol{S}'_n} = \sum_{i=n+1}^{2n} X_{R_i}/n$ and $\rho_{1,2} = \text{cor}(X_{R_1}, X_{R_2})$, then the result of Zimmerman implies that $\text{cor}(X_{\boldsymbol{S}_n}, X_{\boldsymbol{S}'_n}) = n\rho_{1,2}/(1 + (n-1)\rho_{1,2})$. However, $\text{cor}(X_{\boldsymbol{S}_n}, X_{\boldsymbol{S}'_n})$ is one real number, computed as a correlation taken over all possible random test forms, whereas $\rho(\boldsymbol{S}_n)$ is a random variable that can have a different value for each test form, and this is the reliability that a test maker will obtain after computing the reliability with an IRT model for a single test form. The present article will investigate $\rho(\boldsymbol{S}_n)$.

**Example**

Table 1 shows a simple example where it assumed that there are only three items: A, B, and C, with $\varepsilon_A^2 = 3, \varepsilon_B^2 = 6$ an $\varepsilon_C^2 = 9$. Solely for the purpose of this example it is assumed that the same item can be administered multiple times such that the error-score variables are experimentally independent. Alternatively, one can view A, B, and C as different classes of items, each containing many similar items, such that distinct items can be used if the same class is sampled repeatedly. All possible test forms of three items are enumerated in the table. One possible test form displayed in row 9 of Table 1: the randomly selected items in this form are successively A, C, and C, thus $R_1 = A$, $R_2 = C$, and $R_3 = C$, which can also be described as $\boldsymbol{S}_3 = \text{ACC}$. For this test form the resulting item error variances are $\varepsilon^2(R_1) = 3$, $\varepsilon^2(R_2) = 9$, and $\varepsilon^2(R_3) = 9$, yielding an average item error variance $\overline{\varepsilon^2}(\text{ACC}) = 7$. In this example it is furthermore assumed that the true-score variance is $\tau^2 = 1$, and therefore $\rho(\text{ACC}) = 1/(1 + 7/3) = 0.30$. Taking all rows together, we have $\text{Median}(\rho(\boldsymbol{S}_1)) = 0.14$ and $\text{Median}(\rho(\boldsymbol{S}_3)) = 0.33$. Note that these medians agree with the Spearman-Brown formula, since $3 * 0.14/(1 + (3-1) * 0.14) = 0.33$. The approximation is less precise if the expected values are used: $\mathbb{E}(\rho(\boldsymbol{S}_1)) = 0.164$ and $\mathbb{E}(\rho(\boldsymbol{S}_3)) = 0.34$, but $3 * 0.164/(1 + (3-1) * 0.164) = 0.37$.

**Theorem on Reliabilities in Long Random Parallel Test Forms**

In this section the reliabilities will be studied theoretically. The question is whether reliabilities of random parallel tests approximately satisfy the Spearman-Brown formula if the test length goes to infinity. The condition $n \to \infty$ entails that $\rho_n \to 1$, which renders the claim of approximation a tad trivial. Let us therefore transform the reliabilities back to a single scale. For this purpose, define the function

$$SB(x, n) = \frac{nx}{1 + (n-1)x}$$

For fixed $n$, the inverse function for $x$ is $SB^{-1}(x, n) = SB(x, 1/n)$, which is often implicitly used in calculations of lengthening or shortening the test with the Spearman-Brown formula. We will therefore study whether

$$SB\left(\rho(\boldsymbol{S}_n), \frac{1}{n}\right) \text{ converges to a real number for } n \to \infty$$

If so, we may say that *the Spearman-Brown formula holds approximately for large $n$*.

Recall that the section "Notation and general assumptions" described assumptions that hold throughout this article (randomly selected items, conditional independence, finite second moments).

**Theorem**. *Assume that the trues-score variables are bounded by some square integrable random variable $T_{\max}$, that is, $|T_\omega| < T_{\max}$ for all $\omega \in \Omega$ and $\mathbb{E}|T_{\max}| < \infty$. For the reliabilities $\rho(\boldsymbol{S}_n)$, as $n \to \infty$, it holds that*

$$SB\left(\rho(\boldsymbol{S}_n), \frac{1}{n}\right) \to \frac{\mathrm{var}(T_\infty)}{\mathrm{var}(T_\infty) + \mathbb{E}(\varepsilon^2(R_1))}$$

*with probability 1.*

The proof of the theorem is deferred to the appendix, but the basic idea of the proof is that $T_{S_n} \to T_\infty$ by the strong law of large numbers for conditional expectations (Majerek et al., 2005), and $\sum_{i=1}^{n} \varepsilon^2(R_i)/n \to \mathbb{E}(\varepsilon^2(R_1))$ by the ordinary strong law of large numbers.

Note that this theorem does not require that the items are unidimensional in any sense; it suffices to have conditional independence of the error-score variables, and randomly selected items, together with finiteness of the relevant moments. Finiteness of the relevant moments is assured if the observed-score variables are bounded, which is usually the case in real psychometric applications. Furthermore, the theorem pertains to 'true' reliabilities. That is, it assumes that the reliabilities are correctly estimated for each test length, and the theorem does not claim that a similar convergence would also hold for estimates such as Cronbach's alpha, which may underestimate the true reliability (the theorem does not contradict it either).

> **Corollary 1**. *As $n \to \infty$, reliabilities provide consistent estimates of the variance components of a one-facet design as defined by generalizability theory.*

> **Corollary 2**. *For large $n$, the Spearman-Brown formula holds approximately for reliabilities in a one-facet design as defined by generalizability theory..*

A different conclusion can be obtained for small $n$. First, note that $\mathbb{E}(T_{R_i}|\mathcal{F}) = T_\infty$ for all $i \in \mathbb{N}$, which implies $\mathbb{E}(T_{S_n}|\mathcal{F}) = T_\infty$, which is the universe-score variable. Now consider the true-score variance over all test forms of length $n$, which is $\text{var}(T_{S_n})$. By the variance decomposition formula, $\text{var}(T_{S_n}) = \text{var}(\mathbb{E}(T_{S_n}|\mathcal{F})) + \mathbb{E}(\text{var}(T_{S_n}|\mathcal{F})) = \text{var}(T_\infty) + \mathbb{E}(\text{var}(T_{S_n}|\mathcal{F}))$. Therefore

$$\text{var}(T_{S_n}) \geq \text{var}(T_\infty)$$

In other words, the finite test true-score variance is at least as great as the universe-score variance. This is comparable to the conclusion of Lord and Novick (1968, p. 186) that "the generic true-score variance is smaller than the specific true-score variance averaged over tests, except in the unusual case where all test forms are strictly parallel". The present statement is different from this in that it applies to a different true-score concept. Another similar observation is that the reliability with respect to true-scores that include specific factors is larger than the reliability with respect to true scores with respect to only a general factor (Zinbarg et al., 2005).

The test true score variance is different for each test form, however, and a second decomposition is $\text{var}(T_{S_n}) = \text{var}(\mathbb{E}(T_{S_n}|S_n)) + \mathbb{E}(\text{var}(T_{S_n}|S_n))$. The first term equals the variance of the test form means, and the second term, $\text{var}(T_{S_n}|S_n)$, is the true-score variance in test form $S_n$. If we assume for a moment that the item scores are centred, then $\text{var}(\mathbb{E}(T_{S_n}|S_n)) = 0$, and then we obtain

$$\mathbb{E}(\text{var}(T_{S_n}|S_n)) \geq \text{var}(T_\infty)$$

In other words, for centred items the average test-form-specific true-score variance is greater than or equal to the universe-score variance. This inequality does not directly imply that the average reliability is greater than or equal to the generalizability, because that depends also on the error variances. But it should not come as a surprise if this is observed in some settings during the simulations that will be discussed now.

### Simulation Study 1: Unidimensional versus Multidimensional

In this section it will be investigated how well the approximation established in the theorem of the previous section holds in relatively short tests. To this end, Monte Carlo simulations generated item pools of 1000 items that satisfied the multidimensional 2-parameter logistic (2PL) model with up to five dimensions, $\Theta = (\Theta_1, \dots, \Theta_5)$, where each item loaded on precisely one dimension, denoted as $\dim(\omega)$. The probability of a positive response is then

$$P(X_\omega = 1|\Theta) = 1/(1 + \exp(-Da_\omega(\Theta_{\dim(\omega)} - b_\omega))$$

with $D = 1.7$. The item parameters $a_\omega$ and $b_\omega$ were generated by a 4-parameter beta distribution with hyperparameters are $\alpha, \beta$, minimum and maximum). The item pools were designed with the following characteristics:

1. Number of dimensions. Item pools of 1, 2, or 5 dimensions were studied. Each item loaded on only one dimension. Each dimension in the pool was equally probably, that is, each dimension is expected to have approximately the same number of items.

2. Maximum discrimination parameter. The discrimination parameter $a_\omega$ was sampled from a beta distribution with minimum 0 and maximum either $a_{\max} = 2$ or $a_{\max} = 5$. Although the maximum 5 is rarely seen in psychological tests, it can be obtained in health care applications (Hays et al., 2000, p. 4; Yang & Kao, 2015, p. ).

3. Shape of the distribution of discrimination parameter. The $\alpha$ and $\beta$ hyperparameters of the beta distribution of the discrimination parameter $a_\omega$ were set such that the distribution was unimodal with $\alpha + \beta = 12$ or (reverse) J-shaped with $\alpha + \beta = 2$.

4. Mean discrimination parameter. The $\alpha$ and $\beta$ hyperparameters of the beta distribution of the discrimination parameter were set such that the mean of $a_\omega$ could be 0.83, 2.5, or 4.17 if $a_{\max} = 5$, or 0.33, 1.0 or 1.67 if $a_{\max} = 2$.

5. Mean difficulty parameter. The difficulty parameters $b_\omega$ were drawn from a beta distribution with minimum -2 and maximum 2, with an unimodal distribution ($\alpha + \beta = 4$) having mean -1, 0, or 1.

For each value of the number of dimensions (1, 2,or 5) and each value of the maximum discrimination parameter (2 or 5) there were 18 'cases' characterized by the distributions of the discrimination and difficulty parameters (6 possible distributions of $a_\omega$ × 3 possible distributions of $b_\omega$). The online Supplementary Material lists the hyperparameters for all cases. For each case, an item pool of 1000 items was generated, and this pool was used to generate 1000 random test forms of 50 items, in steps of 5 items. At each step the reliability was computed for each test form by numerical integration, assuming a standard normal distribution of the latent ability on each dimension, where the dimensions were independent of each other.

Figures 1 – 3 show how the mean of the reliabilities $\rho(S_n)$ and mean of the rescaled reliabilities $SB\left(\rho(S_n), \frac{1}{n}\right)$ depend on the test length in all cases with $a_{\max} = 2$. The left panels show how the mean reliabilities increase with test length: they follow largely the pattern that is to be expected if the Spearman-Brown formula holds. The right hand panels give a more detailed account via the mean rescaled reliabilities, which should stabilize according to theorem 1. For the one-dimensional tests of Figure 1 the mean rescaled reliabilities are indeed stable, in the sense that each line is approximately horizontal. A Friedman rank test revealed that differences were significant in 10 out of 18 cases, but these effects were small: over all 18 cases, the largest absolute deviation between means of rescaled reliabilities from the same case with different test lengths was 0.0098, which occurred in a case with mean rescaled reliability 0.38. For the two-dimensional tests of Figure 2, however, the mean rescaled reliabilities decrease gradually in the first 20 to 30 items, which implies that these reliabilities have a positive bias for short tests. For the five-dimensional tests of Figure 3 the positive bias is even more pronounced. The rescaled reliabilities are also lower here than in the unidimensional cases, and one may wonder whether this causes the bias. But Figure 4 shows the rescaled reliabilities in the five-dimensional cases with $a_{\max} = 5$; these have about the same magnitude as in the unidimensional cases with $a_{\max} = 2$, and yet there is a clear bias in the five-dimensional case and not in the unidimensional case. The other results with $a_{\max} = 5$ are essentially the same as with $a_{\max} = 2$, and are not displayed.

A related question is whether within each case the mean reliabilities at different test lengths can be predicted from each other. In this setting, lengthening and shortening of the test are associated with different knowledge states. If the investigator has a long test to begin with, and the reliability of this test is known, then the item parameters could usually also be estimated and from this the investigator can calculate what the reliability of any shortened version will be (e.g., Raborn,

Leite & Marcoulides, 2020). On other hand, if the investigator has a short test to begin with, the additional items would often not yet exist, and the Spearman-Brown formula would truly yield a prediction that might be wrong. In this setting, shortening the test requires backward prediction, whereas lengthening the test requires forward prediction.

For each of the mean reliabilities shown in Figures 1 – 3 is was computed how well it can be predicted forwardly from the mean reliabilities with smaller test length, and how well it can be predicted backwardly from the mean reliabilities with larger test lengths. The maximum absolute errors in prediction are shown in Figure 5 as a function of the number of dimensions. The figure shows that the error size depends strongly on the number of dimensions. In the unidimensional cases the errors are less than or equal to 0.02, whereas the maximum errors were between 0.03 and 0.12 in the five-dimensional cases. The explained variances were estimated with a two-factor ANOVA. The number of dimensions explained 84% of the variance in maximum backward errors, while the mean discrimination parameter and its interaction with the number of dimensions explained 2% and 11%, respectively. Together these two factors explained 98% of the variance in maximum backward errors. For the maximum forward errors these percentages were 62% (number of dimensions), 22% (mean discrimination parameter), 8% (interaction), and 92% (together). In the unidimensional cases the largest maximum errors occurred in the cases where the discrimination parameters had a low mean (0.17) with a J-shaped distribution. In the other unidimensional cases the maximum errors were at most 0.006, both in backward and forward prediction.

The conclusion from this section is that we have to reject hypothesis that the Spearman-Brown formula generally yields accurate predictions for short multidimensional tests, but that we cannot yet reject the hypothesis that it yields fairly accurate predictions for short unidimensional tests. Therefore we will focus on unidimensional tests in the second simulation study, and put this hypothesis further to the test.

**Simulation Study 2: Unidimensional With Binary or Irregular Item Parameters**

In this section it will be tested whether the Spearman-Brown formula yields accurate predictions of mean reliabilities in unidimensional tests under the 2PL model. In the previous section the item parameters were drawn from beta-distributions, but the present section will use more irregular distributions in the hope that this may reveal violations of the hypothesis. The following distribution types were studied.

1. Binary item parameters. In order to create extreme situations, cases with $a_\omega \in \{0.5, 2.0\}$ and $b_\omega \in \{-1.7, 1.7\}$ were generated. The probabilities of $a_\omega = 0.5$ were 0.1, 0.3, 0.5, 0.7, or 0.9, and the probabilities of $b_\omega = -1.7$ were 0.1, 0.5, and 0.9. This created 15 cases (5 values of $P(a_\omega = 0.5)$ by 3 values of $P(b_\omega = -1.7)$). An item pool of 1000 items was created for each case.

2. Same, but with $b_\omega \in \{0, 1.7\}$.

3. Irregular distributions of $a_\omega$ and $b_\omega$. This was done by creating 100 cases of small item pools of 10 items each, with of $a_\omega$ and $b_\omega$ drawn from uniform distributions on $[0.5, 2.0]$ and $[-2.0, 2.0]$, respectively. In each item pool the distribution will be irregular because of the small pool size. Moreover, the two item parameters can be correlated within a small item pool.

4. Reported item parameters from the literature, two cases:
    a. Hays et al. (2000, their table 4: 11 items) and
    b. Pedraza et al. (2011, their table 2: 60 items);

In each case a 1000 random test versions of 50 items were created, and the reliabilities were computed while these test versions were growing from 10 to 50 items, in steps of 5 items. The maximum error of prediction, where mean reliabilities associated with different test lengths were predicted from each other, was computed for backward and forward situations separately. The rescaled reliabilities were computed for each of the 9 test lengths and each case, and the maximum absolute difference was computed in each case.

The results are displayed in Table 2. For the prediction of mean reliabilities, the maximum error over all cases was 0.012. For the rescaled mean reliabilities, the maximum error over all cases was 0.017. These error margins are very well acceptable.

### Simulation Study 3: Standard Deviations of the Reliabilities

The previous sections considered how well mean reliabilities could be predicted from each other. In practical situations, however, the test maker will often not have many test versions, but rather a single test version. Even if the test maker knows the correct reliability of this test version (that is, based on a very large subject sample), the test version's reliability might be different from the mean reliability of all test versions of the same length. In order to get an impression of the magnitude of this variation, the standard deviation of the reliabilities was computed in each unidimensional case and each test length of the previous sections. Figure 6 shows boxplots of the standard deviations. As could be expected, the standard deviations tend to decrease within each distribution type as the test length increases. Aggregated over all case the medians of the standard deviations decrease from 0.034 with $n = 10$ to 0.0043 with $n = 50$. The highest standard deviation was 0.131 and the 90$^{th}$ percentile of the standard deviations decreased from 0.074 for $n = 10$ to 0.031 for $n = 50$.

### Discussion

We have investigated to which extent the Spearman-Brown formula can be used to predict the reliability outcomes of changing test length for non-parallel items. It was shown that in a one-facet universe with randomly sampled items with conditionally independent error-scores, the reliabilities are approximately equal to the generalizability for long tests. This holds regardless of the dimensionality of the items. For short tests we conducted simulations where random test versions of items with known item parameters of a 2PL model were drawn, after which the IRT model-based reliabilities were computed with numerical integration. The reliabilities had a substantial positive bias

SPEARMAN-BROWN FORMULA AND RANDOM TEST FORMS                                        20for multidimensional items in short tests. For unidimensional items the reliabilities were almost unbiased, and the mean reliabilities with different test lengths could be predicted from each other via the Spearman-Brown formula. However, a single short test version can have a reliability that deviates substantially from the mean reliability of test versions of the same length from the same item pool, although most cases that were considered had standard deviation less than 0.034 even with $n = 10$.

Returning to the question in the introduction, whether the accuracy of the Spearman-Brown formula depends on the dimensionality of the items or that random sampling of items is sufficient, the answer is: this depends on the test length. The dimensionality does not matter if the test is sufficiently long, as was established in Theorem 1, but for short tests dimensionality is important, as demonstrated in simulation studies.

The conclusion is that the Spearman-Brown formula can reasonably be used to predict the reliability after changing the test length with randomly sampled items, provided that the initial test length is long or the test items are 2PL-unidimensional. In forward prediction from a short test to a long test there is a chance that the reliability of the short test version is relatively far away from the mean reliability, which will contaminate the prediction. However, for 2PL-unidimensional item pools the bias was negligible (despite being significant), and therefore prediction with the Spearman-Brown formula can be viewed as an educated guess, even though the reliability cannot be predicted with certainty - as would be the case with parallel items. For multidimensional item pools on the other hand, there was a clear, non-negligible positive bias in the reliability of short tests. As a consequence, test makers who use the Spearman-Brown formula in multidimensional cases to forwardly predict the reliability of a longer test, will easily be too optimistic.

# Appendix

This appendix contains the proof of the Theorem in the main text. A lemma is first stated and proved, and after that the theorem is stated and proved. Recall that the section "Notation and general assumptions" described assumptions that hold throughout this article (randomly selected items, conditional independence, finite second moments).

Let

$$\overline{\varepsilon^2}(\boldsymbol{S}_n) = \sum_{i=1}^n \varepsilon^2(R_i)/n$$

This is a mean of $n$ i.i.d. variables $\varepsilon^2(R_i)$.

**Lemma 1**. $\mathrm{var}(E_{\boldsymbol{S}_n}|\boldsymbol{S}_n) = \overline{\varepsilon^2}(\boldsymbol{S}_n)/n$ and $\mathbb{E}\left(\overline{\varepsilon^2}(\boldsymbol{S}_n)\right) = \mathbb{E}(\varepsilon^2(R_1))$.

**Proof of Lemma 1.** The $R_i$ are independent, and therefore $\mathrm{var}(E_{R_i}|\boldsymbol{S}_n) = \mathrm{var}(E_{R_i}|R_i) = \varepsilon^2(R_i)$. Then

$$\mathrm{var}(E_{\boldsymbol{S}_n}|\boldsymbol{S}_n) = \sum_{i=1}^n \frac{\mathrm{var}(E_{R_i}|\boldsymbol{S}_n)}{n^2} = \sum_{i=1}^n \frac{\varepsilon^2(R_i)}{n^2} = \overline{\varepsilon^2}(\boldsymbol{S}_n)/n$$

Furthermore, since the $R_i$ are i.i.d, $\mathbb{E}(\varepsilon^2(R_i)) = \mathbb{E}(\varepsilon^2(R_1))$, and therefore

$$\mathbb{E}\left(\overline{\varepsilon^2}(\boldsymbol{S}_n)\right) = \mathbb{E}\left(\sum_{i=1}^n \varepsilon^2(R_i)/n\right) = \frac{\sum_{i=1}^n \mathbb{E}(\varepsilon^2(R_i))}{n} = \mathbb{E}(\varepsilon^2(R_1)) \blacksquare$$

**Theorem 1**. *Assume that the trues-score variables are bounded by some square integrable random variable $T_{\max}$, that is, $|T_\omega| < T_{\max}$ for all $\omega \in \Omega$ and $\mathbb{E}|T_{\max}| < \infty$. For the IRT-based reliabilities $\rho(\boldsymbol{S}_n)$, as $n \to \infty$,*

$$SB\left(\rho(\boldsymbol{S}_n), \frac{1}{n}\right) \to \frac{\mathrm{var}(T_\infty)}{\mathrm{var}(T_\infty) + \mathbb{E}(\varepsilon^2(R_1))}$$

*with probability 1*

**Proof of theorem 1.** By the definition of $\rho(\boldsymbol{S}_n)$ and Lemma 1,

$$\rho(\boldsymbol{S}_n) = \frac{\operatorname{var}(T_{\boldsymbol{S}_n}|\boldsymbol{S}_n)}{\operatorname{var}(T_{\boldsymbol{S}_n}|\boldsymbol{S}_n) + \overline{\varepsilon^2}(\boldsymbol{S}_n)/n}$$

which implies

$$SB\left(\rho(\boldsymbol{S}_n), \frac{1}{n}\right) = \frac{\operatorname{var}(T_{\boldsymbol{S}_n}|\boldsymbol{S}_n)}{\operatorname{var}(T_{\boldsymbol{S}_n}|\boldsymbol{S}_n) + \overline{\varepsilon^2}(\boldsymbol{S}_n)}$$

Consider first convergence of the true score variance. Note that the $T_{R_i}$ are exchangeable. Since it is assumed that $|\mathbb{E}(T_{R_i})| < \infty$, one can define $T_\infty := \mathbb{E}(T_{R_1}|\mathcal{F})$ and show that the $T_{R_i}$ are i.i.d. given $\mathcal{F}$ with lemma 4.1 of Dawid, 1979. By the strong law of large numbers for conditional expectations (Majerek et al., 2005, theorem 4.2), $T_{\boldsymbol{S}_n} \to T_\infty$ with probability 1.

Write $\boldsymbol{S}_\infty = (R_1, R_2, \dots)$. Since $\boldsymbol{S}_n$ and $(R_{n+1}, R_{n+2}, \dots)$ are independent, $\operatorname{var}(T_{\boldsymbol{S}_n}|\boldsymbol{S}_n) = \operatorname{var}(T_{\boldsymbol{S}_n}|\boldsymbol{S}_\infty)$. We have already established that $T_{\boldsymbol{S}_n} \to T_\infty$ with probability 1, and therefore $\operatorname{var}(T_{\boldsymbol{S}_n}|\boldsymbol{S}_\infty) \to \operatorname{var}(T_\infty|\boldsymbol{S}_\infty)$ with probability 1 by the dominated convergence theorem for conditional expectations (e.g., Billingsley, 1986, Th.34.2.v), using the hypothesis that the $|T_\omega|$ are dominated. However, $T_\infty$ is defined solely in terms of $\mathcal{F}$, which is independent of $\boldsymbol{S}_\infty$; therefore $\operatorname{var}(T_\infty|\boldsymbol{S}_\infty) = \operatorname{var}(T_\infty)$. In sum, $\operatorname{var}(T_{\boldsymbol{S}_n}|\boldsymbol{S}_n) \to \operatorname{var}(T_\infty)$.

Next, consider convergence of the error variance. By the strong law of large numbers (which may be applied because the $R_i$ are independent and $\operatorname{var}(\varepsilon^2(R_i)) < \infty$),

$$\overline{\varepsilon^2}(\boldsymbol{S}_n) \to \mathbb{E}(\varepsilon^2(R_1))$$

therefore (note that it was assumed that $\varepsilon^2(R_1) > 0$)

$$SB\left(\rho(\boldsymbol{S}_n), \frac{1}{n}\right) \to \frac{\operatorname{var}(T_\infty)}{\operatorname{var}(T_\infty) + \mathbb{E}(\varepsilon^2(R_1))}$$

∎

**Table 1**

*Example with Enumeration of All Possible Test Forms of Three Items and Their Error Variances.*

| $R_1$ | $R_2$ | $R_3$ | $S_3$ | $\varepsilon^2(R_1)$ | $\varepsilon^2(R_2)$ | $\varepsilon^2(R_3)$ | $\overline{\varepsilon^2}(S_3)$ | $\rho(S_3)$ | $\rho(S_1)$ |
|---|---|---|---|---|---|---|---|---|---|
| A | A | A | AAA | 3 | 3 | 3 | 3 | 0.50 | 0.25 |
| A | A | B | AAB | 3 | 3 | 6 | 4 | 0.43 | 0.25 |
| A | A | C | AAC | 3 | 3 | 9 | 5 | 0.38 | 0.25 |
| A | B | A | ABA | 3 | 6 | 3 | 4 | 0.43 | 0.25 |
| A | B | B | ABB | 3 | 6 | 6 | 5 | 0.38 | 0.25 |
| A | B | C | ABC | 3 | 6 | 9 | 6 | 0.33 | 0.25 |
| A | C | A | ACA | 3 | 9 | 3 | 5 | 0.38 | 0.25 |
| A | C | B | ACB | 3 | 9 | 6 | 6 | 0.33 | 0.25 |
| A | C | C | ACC | 3 | 9 | 9 | 7 | 0.30 | 0.25 |
| B | A | A | BAA | 6 | 3 | 3 | 4 | 0.43 | 0.14 |
| B | A | B | BAB | 6 | 3 | 6 | 5 | 0.38 | 0.14 |
| B | A | C | BAC | 6 | 3 | 9 | 6 | 0.33 | 0.14 |
| B | B | A | BBA | 6 | 6 | 3 | 5 | 0.38 | 0.14 |
| B | B | B | BBB | 6 | 6 | 6 | 6 | 0.33 | 0.14 |
| B | B | C | BBC | 6 | 6 | 9 | 7 | 0.30 | 0.14 |
| B | C | A | BCA | 6 | 9 | 3 | 6 | 0.33 | 0.14 |
| B | C | B | BCB | 6 | 9 | 6 | 7 | 0.30 | 0.14 |
| B | C | C | BCC | 6 | 9 | 9 | 8 | 0.27 | 0.14 |
| C | A | A | CAA | 9 | 3 | 3 | 5 | 0.38 | 0.10 |
| C | A | B | CAB | 9 | 3 | 6 | 6 | 0.33 | 0.10 |
| C | A | C | CAC | 9 | 3 | 9 | 7 | 0.30 | 0.10 |
| C | B | A | CBA | 9 | 6 | 3 | 6 | 0.33 | 0.10 |
| C | B | B | CBB | 9 | 6 | 6 | 7 | 0.30 | 0.10 |
| C | B | C | CBC | 9 | 6 | 9 | 8 | 0.27 | 0.10 |
| C | C | A | CCA | 9 | 9 | 3 | 7 | 0.30 | 0.10 |
| C | C | B | CCB | 9 | 9 | 6 | 8 | 0.27 | 0.10 |
| C | C | C | CCC | 9 | 9 | 9 | 9 | 0.25 | 0.10 |
| | | | | | | | Median: | 0.33 | 0.14 |
| | | | | | | | Mean: | 0.34 | 0.16 |

Note. The items are selected from classes A, B, and C with item error variances 3, 6, and 9, respectively. $R_i$ is the *i*-th item, $S_3$ is the test form, $\varepsilon^2$ is the item error variance, $\overline{\varepsilon^2}$ is the test forms mean error variance, and $\rho(S_3)$ is the reliability of the test form. $\rho(S_1)$ is the reliability of the test form consisting only of the first item (i.e., $S_1 = R_1$). In the computation of $\rho(S_3)$ and $\rho(S_1)$ it is assumed that the true-score variance is $\tau^2 = 1$.

**Table 2**

*Maximum Errors of Simulation Study 2.*

| Distribution type | Maximum error in backward or forward prediction of mean reliabilities | Maximum absolute difference of rescaled mean reliabilities |
| --- | --- | --- |
| 1 | 0.012 | 0.017 |
| 2 | 0.005 | 0.009 |
| 3 | 0.006 | 0.013 |
| 4a | 0.002 | 0.003 |
| 4b | 0.0004 | 0.003 |

**Figure 1**

*Mean reliabilities in Simulation Study 1 for unidimensional cases with $a_{\max} = 2$.*

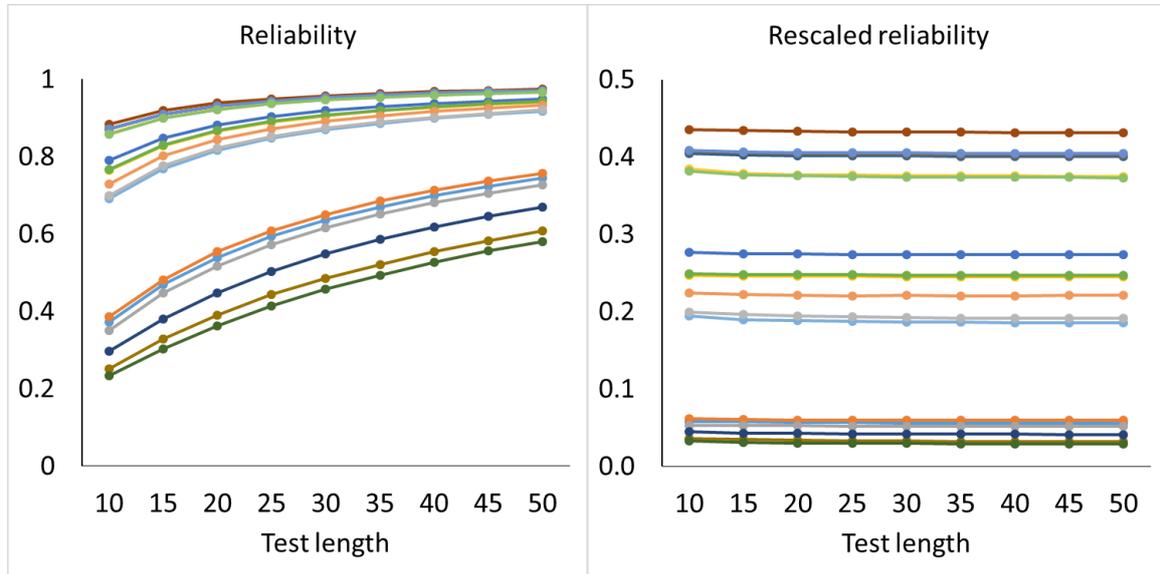

Note. Mean reliability and mean rescaled reliability as a function of test length, in 18 cases of unidimensional models. The cases are represented by different colors. Each point is based on 1000 random test versions.

**Figure 2**

*Mean reliabilities in Simulation Study 1 for two-dimensional cases with $a_{\max} = 2$.*

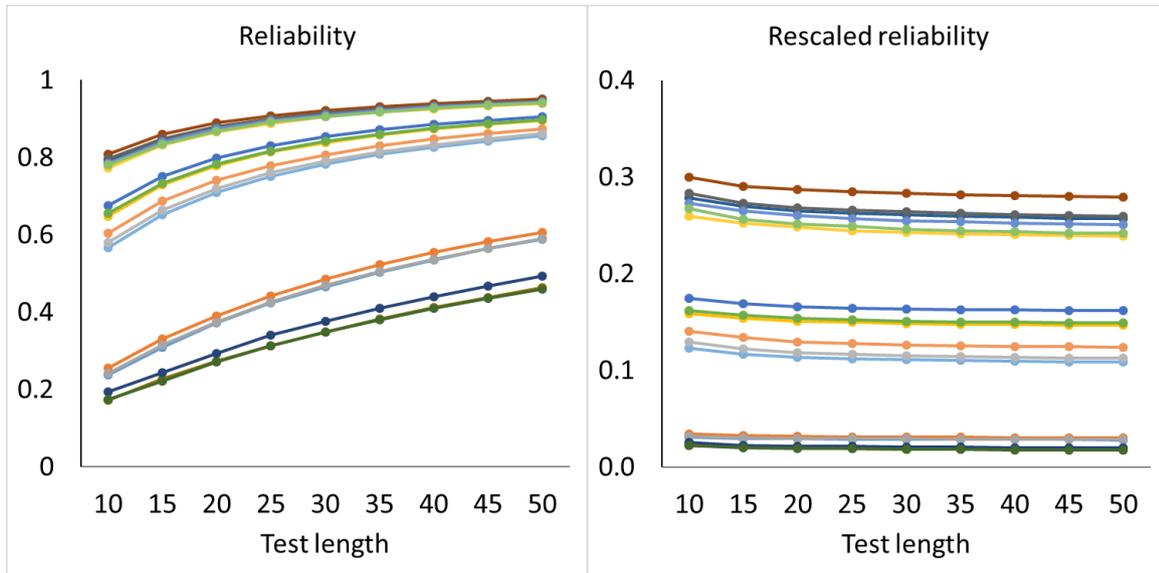

Note. Mean reliability and mean rescaled reliability as a function of test length, in 18 cases of two-dimensional models. The cases are represented by different colors. Each point is based on 1000 random test versions.

**Figure 3**

*Mean reliabilities in Simulation Study 1 for five-dimensional cases with $a_{\max} = 2$.*

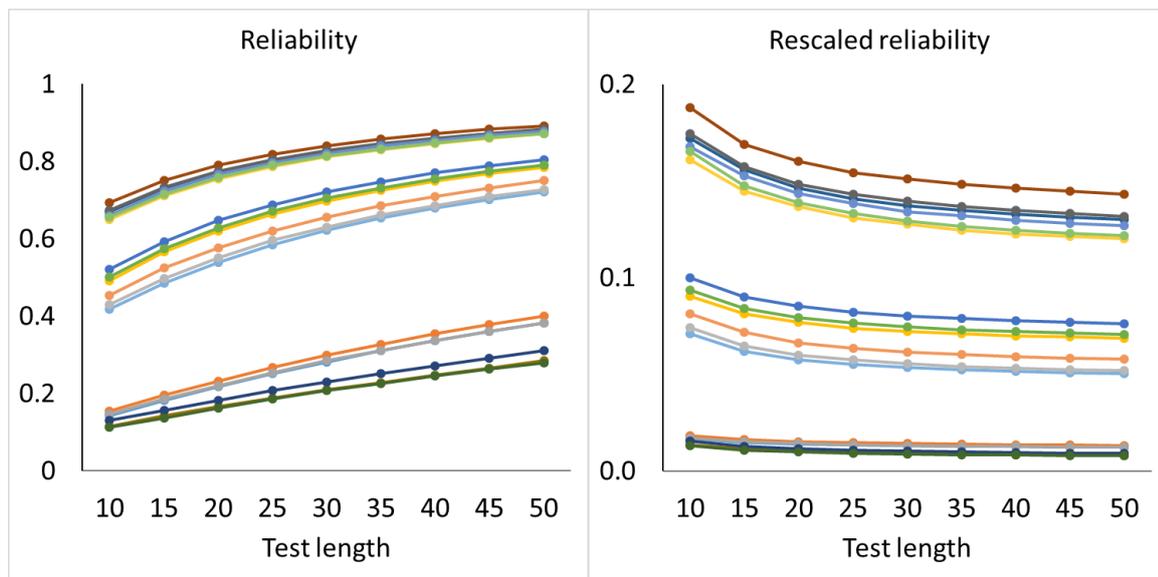

Note. Mean reliability and mean rescaled reliability as a function of test length, in 18 cases of five-dimensional models. The cases are represented by different colors. Each point is based on 1000 random test versions.

**Figure 4**

*Mean reliabilities in Simulation Study 1 for five-dimensional cases with $a_{max} = 5$.*

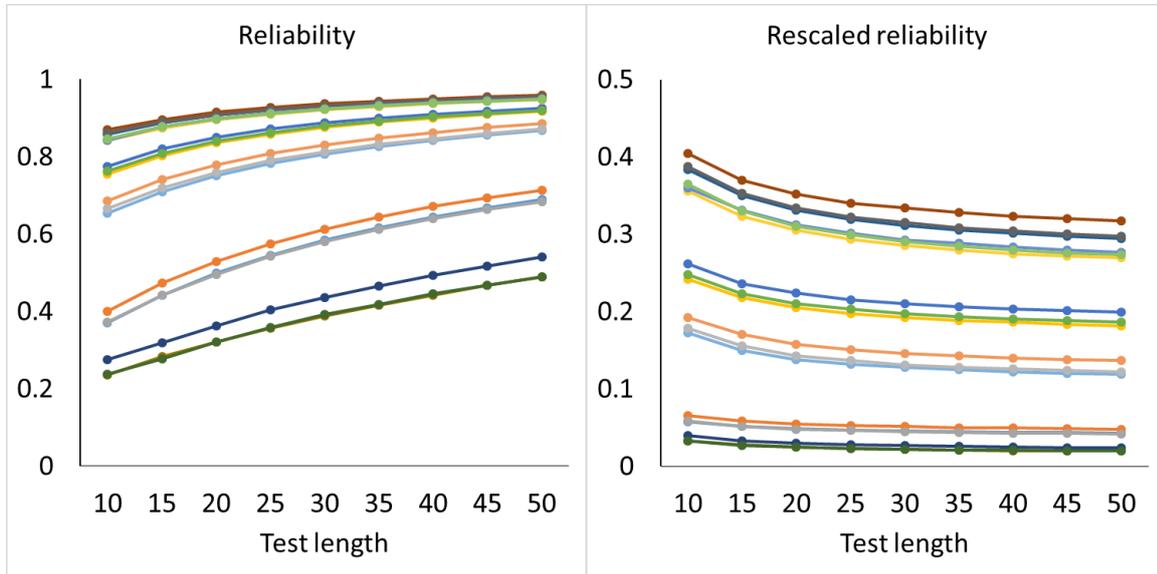

Note. Mean reliability and mean rescaled reliability as a function of test length, in 18 cases of five-dimensional models. The cases are represented by different colors. Each point is based on 1000 random test versions.

**Figure 5**

Boxplots of the maximum errors in prediction of mean reliabilities.

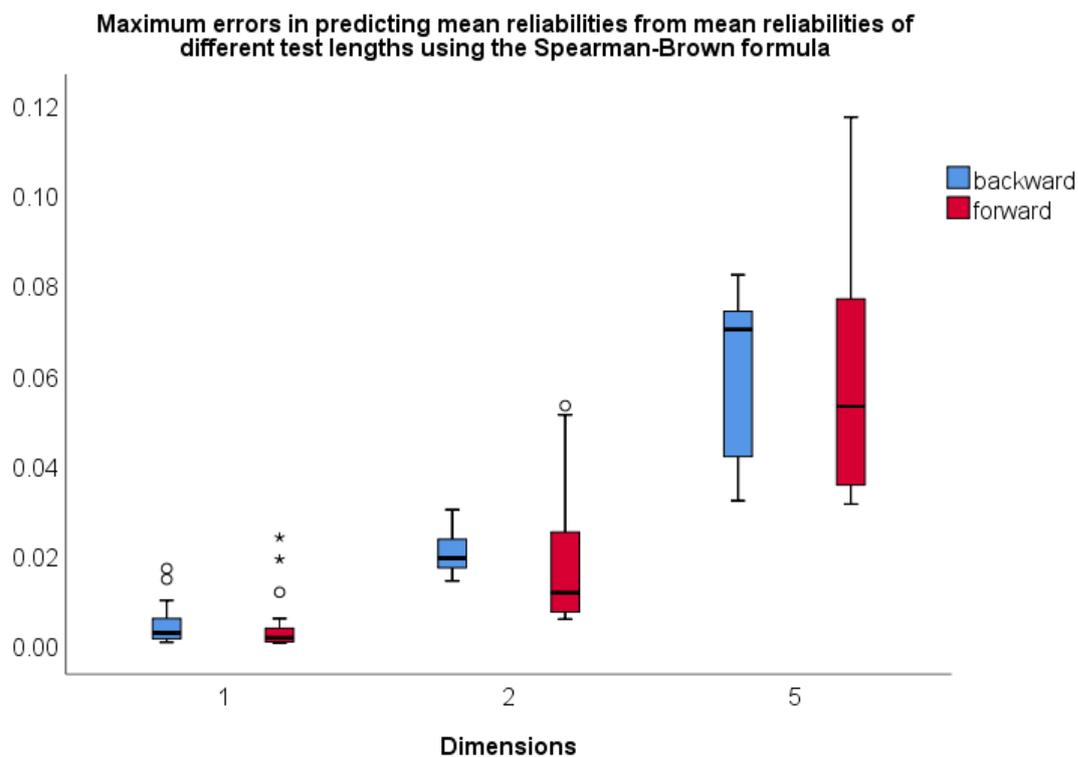

Note. Each boxplot is based on 18 maxima, corresponding to 18 cases with $a_{max} = 2$. Each maximum is based on 36 predictions of a mean reliability of one test length from a mean reliability of another test length, for 9 different test lengths.

**Figure 6**

*Boxplots of the standard deviations of the reliabilities as a function of the distribution type of the parameters and the length of the test versions.*

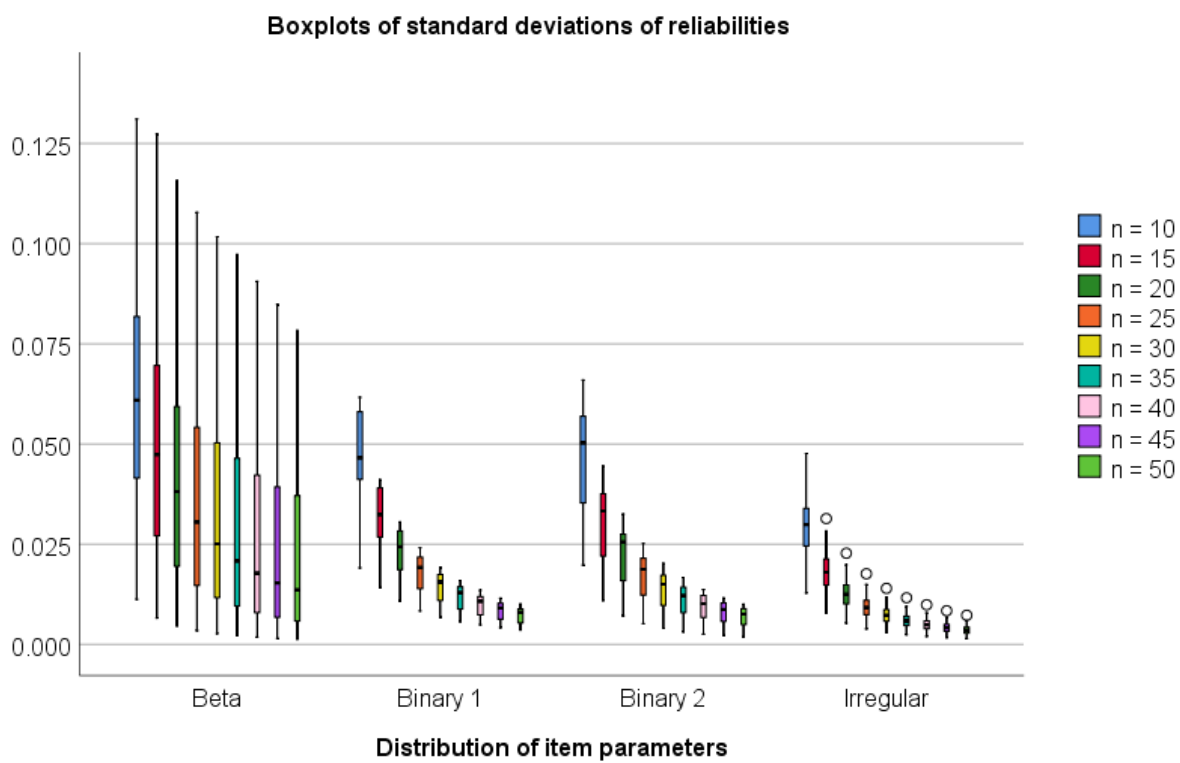

Note. Only unidimensional items with $a_{max} = 2$ were used in this plot. Each boxplot is based on 54 (Beta) or 15 (Binary 1 and Binary 2) or 100 (Irregular) standard deviations, and each standard deviation is based on 1000 reliabilities of random test forms drawn from the same item pool.